# A new upper bound on the query complexity for testing generalized Reed-Muller codes

Noga Ron-Zewi [*]   Madhu Sudan[†]

October 12, 2018

## Abstract

Over a finite field $\mathbb{F}_q$ the $(n, d, q)$-Reed-Muller code is the code given by evaluations of $n$-variate polynomials of total degree at most $d$ on all points (of $\mathbb{F}_q^n$). The task of testing if a function $f : \mathbb{F}_q^n \to \mathbb{F}_q$ is close to a codeword of an $(n, d, q)$-Reed-Muller code has been of central interest in complexity theory and property testing. The query complexity of this task is the minimal number of queries that a tester can make (minimum over all testers of the maximum number of queries over all random choices) while accepting all Reed-Muller codewords and rejecting words that are $\delta$-far from the code with probability $\Omega(\delta)$. (In this work we allow the constant in the $\Omega$ to depend on $d$.)

For codes over a prime field $\mathbb{F}_q$ the optimal query complexity is well-known and known to be $\Theta(q^{\lceil (d+1)/(q-1) \rceil})$, and the test consists of testing if $f$ is a degree $d$ polynomial on a randomly chosen ($\lceil (d + 1)/(q - 1) \rceil$)-dimensional affine subspace of $\mathbb{F}_q^n$. If $q$ is not a prime, then the above quantity remains a lower bound, whereas the previously known upper bound grows to $O(q^{\lceil (d+1)/(q-q/p)\rceil})$ where $p$ is the characteristic of the field $\mathbb{F}_q$. In this work we give a new upper bound of $(cq)^{(d+1)/q}$ on the query complexity, where $c$ is a universal constant. Thus for every $p$ and sufficiently large constant $q$ this bound improves over the previously known bound by a polynomial factor, as we let $d \to \infty$.

In the process we also give new upper bounds on the "spanning weight" of the dual of the Reed-Muller code (which is also a Reed-Muller code). The spanning weight of a code is the smallest integer $w$ such that codewords of Hamming weight at most $w$ span the code. The main technical contribution of this work is the design of tests that test a function by *not* querying its value on an entire subspace of the space, but rather on a carefully chosen (algebraically nice) subset of the points from low-dimensional subspaces.

[*]Department of Computer Science, Technion, Haifa. nogaz@cs.technion.ac.il. Research was conducted while the author was an intern at Microsoft Research New-England, Cambridge, MA, and supported by the Israel Ministry of Science and Technology.

[†]Microsoft Research New-England, Cambridge, MA. madhu@mit.edu

# 1 Introduction

In this work we present new upper bounds on the query complexity of testing Reed-Muller codes, the codes obtained by evaluations of multivariate low-degree polynomials, over general fields. In the process we also give new upper bounds on the spanning weight of Reed-Muller codes. We explain these terms and our results below.

We start with the definition of Reed-Muller codes. Let $\mathbb{F}_q$ denote the finite field on $q$ elements. Throughout we will let $q = p^s$ for prime $p$ and integer $s$. The Reed-Muller codes have two parameters in addtion to the field size, namely the degree $d$ and number of variables $n$. The $(n, d, q)$-Reed-Muller code $\text{RM}[n, d, q]$ is the set of functions from $\mathbb{F}_q^n$ to $\mathbb{F}_q$ that are evaluations of $n$-variate polynomials of total degree at most $d$.

## 1.1 Testing Reed-Muller Codes

We define the notion of testing the "Reed-Muller" property as a special case of property testing. We let $\{\mathbb{F}_q^n \to \mathbb{F}_q\}$ denote the set of all functions mapping $\mathbb{F}_q^n$ to $\mathbb{F}_q$. A property $\mathcal{F}$ is simply a subset of such functions. For $f, g : \mathbb{F}_q^n \to \mathbb{F}_q$ we say the distance between them $\delta(f, g)$ is the fraction of points of $\mathbb{F}_q^n$ where they disagree. We let $\delta(f, \mathcal{F})$ denote the minimum distance between $f$ and a function in $\mathcal{F}$. We say $f$ is $\delta$-close to $\mathcal{F}$ if $\delta(f, \mathcal{F}) \leq \delta$ and $\delta$-far otherwise.

A $(k, \epsilon)$-tester for the property $\mathcal{F} \subseteq \{\mathbb{F}_q^n \to \mathbb{F}_q\}$ is a randomized algorithm that makes at most $k$ queries to an oracle for a function $f : \mathbb{F}_q^n \to \mathbb{F}_q$ and accepts if $f \in \mathcal{F}$ and rejects $f \notin \mathcal{F}$ with probability at least $\epsilon \delta(f, \mathcal{F})$.

For fixed $d$ and $q$, we consider *query complexity* of testing the property of being a degree $d$ multivariate polynomial over $\mathbb{F}_q$. Specifically, the query complexity $k = k(d, q)$, is the minimum integer such that there exists an $\epsilon$ such that for all $n$ there is a $(k, \epsilon)$-tester for the $\text{RM}[n, d, q]$ property. (So the error $\epsilon$ of the tester is allowed to depend on $q$ and $d$, but not on $n$.)

The query complexity of low-degree testing is a well-studied question and has played a role in many results in computational complexity including in the PCP theorem ([ALM+98] and subsequent works), and in the works of Viola and Wigderson [VW08] and Barak et al. [BGH+11]. Many of these results depend not only on a tight analysis of $k(d, q)$ but also a tight analysis of the parameter $\epsilon$, but in this work we only focus on the first quantity. Below we describe what was known about these quantities.

For the case when $d$ is (sufficiently) smaller than the field size, the works of Rubinfeld and Sudan [RS96] and Friedl and Sudan [FS95] show that $k(d, q) = d + 2$ (provided $d < q - q/p$). For the case when $q = 2$ and $d$ is arbitrary, this quantity was analyzed in the work of Alon et al [AKK+05] who show that $k(d, 2) = 2^{d+1}$ (exactly). Jutla et al [JPRZ09] and Kaufman and Ron [KR06] explored this question for general $q$ and $d$ (the former only considered prime $q$) and showed that $k(d, q) \leq q^{\lceil (d+1)/(q-q/p) \rceil}$. In [KR06] it is also shown that the bound is tight (to within a factor of $q$) if $q$ is a prime. However for the non-prime case the only known lower bound on the query complexity was $k(d, q) \geq q^{(d+1)/(q-1)}$ (which is roughly the upper bound raised to the power of $(p-1)/p$). (In the following sections we describe the conceptual reason for this gap in knowledge.)

In this work we give a new upper bound on $k(d, q)$ which is closer to the lower bound when $p$ is a constant and $d$ and $q$ are going to infinity. We state our main theorem below.

**Theorem 1.1** (Main). *Let $q = p^s$ for prime $p$ and positive integer $s$. Then there exists a constant $c_q \leq 3q^4$ such that for every $d$ and $n$, the Reed-Muller code $\text{RM}[n, d, q]$ has a $(k, \Omega(1/k^2))$-local tester, for $k = k(d, q) \leq c_q \cdot \left(2^{p-1} + p - 1\right)^{(d+1)/(q(p-1))} q^{(d+1)/q}$. In particular $k(d, q) \leq 3q^4 \cdot (3q)^{(d+1)/q}$.*

We note that when $p$ goes to infinity the bound on $k(d,q)$ tends to $c_q \cdot (3q)^{(d+1)/q}$. We also note that the constant $c_q$ is not optimized in our proofs and it seems quite plausible that it can be improved using more careful analysis. The more serious factor (especially when one considers a constant $q$ and $d \to \infty$) is the constant factor multiplying $q$ in the base of the exponent. Our techniques do seem to be unable to improve this beyond $(2^{p-1} + p - 1)^{1/(p-1)}$ which is always between 2 and 3 (while the lower bounds suggest a constant which is close to 1).

Theorem 1.1 is proved by proving that the Reed-Muller code $\mathrm{RM}[n, d, q]$ has a "$k$-single-orbit characterization" (a notion we will define later, see Definition 2.2 and Theorem 2.4). This will imply the testing result immediately by a result of Kaufman and Sudan [KS07].

## 1.2 Spanning weight

It is well-known (cf. [BHR05]) that the query complexity of testing a linear code $C$ is lower bounded by the "minimum distance" of its dual, where the minimum distance of a code is the minimum weight of a non-zero codeword. (The weight of a word is simply the number of non-zero coordinates.) Applied to the Reed-Muller code $\mathrm{RM}[n, d, q]$ this suggests a lower bound via the minimum distance of its dual, which also turns out to be a Reed-Muller code. Specifically the dual of $\mathrm{RM}[n, d, q]$ is $\mathrm{RM}[n, n(q-1) - d - 1, q]$. The minimum distance of the latter is well-known and is (roughly) $q^{(d+1)/(q-1)}$ and this leads to the tight analysis of the query complexity of Reed-Muller codes over prime fields.

Over non-prime fields however this bound has not been matched, so one could turn to potentially stronger lower bounds. A natural such bound would be the "spanning weight" of the dual code, namely the minimum weight $w$ such that codewords of the dual of weight at most $w$ span the dual code. It is easy to show that to achieve any positive $\epsilon$ (even going to 0 as $n \to \infty$) a $(k, \epsilon)$-local tester must make at least $w$ queries (on some random choices), where $w$ is the spanning weight of the dual. Somewhat surprisingly, the spanning weight of the Reed-Muller code does not seem well-understood. (Some partial understanding comes from [DK00].). Since for a linear code, the spanning weight of its dual code is a lower bound on the query complexity of the code, our result gives new upper bounds on this spanning weight. Specifically, we have

**Corollary 1.2.** *Let $q = p^s$ for prime $p$ and positive integer $s$. Then there exists a constant $c_q \leq 3q^4$ such that for every $d$ and $n$, the Reed-Muller code $\mathrm{RM}[n, n(q-1) - d - 1, q]$ has a spanning weight of at most $c_q \cdot (2^{p-1} + p - 1)^{(d+1)/(q(p-1))} \cdot q^{(d+1)/q} \leq 3q^4 \cdot (3q)^{(d+1)/q}$.*

## 1.3 Qualitative description and techniques

Our tester differs from previous ones in some qualitative ways. All previously analyzed testers for low-degree testing roughly worked as follows: They picked a large enough dimension $t$ (depending on $q$ and $d$, but not $n$) and verified that the function to be tested was a degree $d$ polynomial on a random $t$-dimensional affine subspace. The final aspect was verified by querying the function on the entire $t$-dimensional space, thus leading to a query complexity of $q^t$. The minimal choice of the dimension $t$ that allows this test to detect functions that are not degree $d$ polynomials with positive probability is termed the "testing dimension" (see, for instance, [HSS11]), and this quantity is well-understood, and equals $t_{q,d} = \lceil (d+1)/(q - q/p) \rceil$.

Any improvement to the query complexity of the test above requires two features: (1) For some choices of the tester's randomness, the set of queried points should span a $t_{q,d}$ dimensional space. (2) For all choices of the tester's randomness, it should make $o(q^{t_{q,d}})$ queries. Finding such a useful subset of $\mathbb{F}_q^n$ turns out to be a non-trivial task. The fortunate occurence that provides the basis for our tester is that such sets of points can indeed be found, and even (in retrospect) systematically.

To illustrate the central idea, consider the setting of $n = 2$, $d = q - 1$ and $q = 2^s$ for some large $s$. While the naive test would query the given function $f : \mathbb{F}_q^2 \to \mathbb{F}_q$ at all $q^2$ points, we wish to query only $O(q)$ points. Our test, for this simple setting is the following: We pick a random affine-transformation $T : \mathbb{F}_q^2 \to \mathbb{F}_q^2$ and test that the function $f \circ T$ has a zero "inner-product" with the function $g : \mathbb{F}_q^2 \to \mathbb{F}_q$ given by $g(x, y) = \frac{1}{y}((x + y)^{q-1} - x^{q-1})$. Here "inner-product" is simply the quantity $\sum_{\alpha, \beta \in \mathbb{F}_q} (f \circ T)(\alpha, \beta) g(\alpha, \beta)$. It can be verified that the function $g$ is zero very often and indeed takes on non-zero values on at most $3q = O(q)$ points in $\mathbb{F}_q^2$. So querying $f(\alpha, \beta)$ at these $O(q)$ points suffices. The more interesting question is: Why is this test complete and sound?

Completeness is also easy to verify. It can be verified, by some simple manipulations that any monomial of the form $x^i y^j$ with $i + j < q$ has a zero inner product with $g$ and by linearity of the test it follows that all polynomials of total degree at most $d$ have a zero inner product with $g$. Since the degree of functions is preserved under affine-transformations, it then follows that $f \circ T$ also has zero inner product with $g$ for every polynomial $f$ of total degree at most $d$.

Finally, we turn to the soundness. Here we appeal to the emerging body of work on affine-invariant linear properties (linear properties that are preserved under affine-transformations), which allows us to focus on very specific monomials and to verify that their inner product with $g$ is non-zero. Specifically, the theory allows us to focus on only the monomials $x^i y^{q-i}$ and for these monomials one can again verify that their inner product with $g$ is non-zero. Using the general methods in the theory of affine-invariant property testing, one can conclude that all polynomials of degree greater than $d$ are rejected with positive probability.

Extending the above result to the general case turns out relatively clean, again using methods from the study of testing of affine-invariant linear properties. The extension to general $n$ is immediate. Extending to other degrees involves some intuitive ways of combining tests, with analysis that get simplified by the emerging theory. These combinations yield the query complexity of roughly $(3q)^{(d+1)/q}$. We however attempt to reduce the constant in front of $q$ in the base of this expression and manage to get an expression that tends to 2 when $p$ goes to infinity. In order to do so we abstract the function $g$ as being the derivative of the function $x^{q-1}$ in direction $y$, and extend it to use iterative derivatives. This yields the best tests we give in the paper.

**Organization** In Section 2 we introduce some of the standard background material from the study of affine-invariant linear properties and use the theory to provide restatements of our problem. In Section 3 we introduce the main novelty of our work, which provides a restricted version of our test while achieving significant savings over standard tests. In Section 4 we build on the test from the previous section and extend it to get a tester for the general case.

## 2 Background and restatement of problem

We start by introducing some of the background material that leads to some reformulations of the main theorem we wish to prove. We first introduce the notions of "constraints" and "(single-orbit) characterizations", which leads to a first reformulation of our main theorem (see Theorem 2.4). We then give some sufficient conditions to recognize such characterizations, and this leads to a second reformulation of our main theorem (see Theorem 2.13).

### 2.1 Single-orbit characterizations

In this section we use the fact that Reed-Muller codes form a "linear, affine-invariant property". We recall these notions first. Given a finite field $\mathbb{F}_q$ a property is a set of functions $\mathcal{F}$ mapping $\mathbb{F}_q^n$ to $\mathbb{F}_q$. The property is said to be *linear* if it is an $\mathbb{F}_q$-vector space, i.e., $\forall f, g \in \mathcal{F}$ and $\alpha \in \mathbb{F}_q$ we have $\alpha f + g \in \mathcal{F}$. The property is said to be *affine-invariant* if it is invariant under affine-transformations of the domain, i.e., $\forall f \in \mathcal{F}$ it is the case that $f \circ T$ is also in $\mathcal{F}$ for every affine-transformation $T : \mathbb{F}_q^n \to \mathbb{F}_q^n$ given by $T(x) = A \cdot x + \beta$ for $A \in \mathbb{F}_q^{n \times n}$, $\beta \in \mathbb{F}_q^n$.[1] It can be easily verified that $\mathrm{RM}[n, d, q]$ is linear and affine-invariant for every $n, d, q$.

The main tool used so far for constructing testers for affine-invarinat linear properties is a structural theorem which shows that every linear affine-invariant property that is $k$-single characterizable is also $k$-locally testable. In order to describe the notion of single-orbit characterizability we start with a couple of definitions.

**Definition 2.1** ($k$-constraint, $k$-characterization). A *$k$-constraint* $C = (\overline{\alpha}, \{\overline{\lambda}_i\}_{i=1}^r)$ on $\{\mathbb{F}_q^n \to \mathbb{F}_q\}$ is given by a vector $\overline{\alpha} = (\alpha_1, \ldots, \alpha_k) \in (\mathbb{F}_q^n)^k$ together with $r$ vectors $\overline{\lambda}_i = (\lambda_{i,1}, \ldots \lambda_{i,k}) \in \mathbb{F}_q^k$ for $1 \leq i \leq r$. We say that the constraint $C$ *accepts* a function $f : \mathbb{F}_{q^n} \to \mathbb{F}_q$ if $\sum_{j=1}^k \lambda_{i,j} f(\alpha_j) = 0$ for all $1 \leq i \leq r$. Otherwise we say that $C$ *rejects* $f$.

Let $\mathcal{F} \subseteq \{\mathbb{F}_{q^n} \to \mathbb{F}_q\}$ be a linear property. A *$k$-characterization* of $\mathcal{F}$ is a collection of $k$-constraints $C_1, \ldots, C_m$ on $\{\mathbb{F}_q^n \to \mathbb{F}_q\}$ such that $f \in \mathcal{F}$ if and only if $C_j$ accepts $f$, for every $j \in \{1, \ldots, m\}$.

It is well-known [BHR05] that every $k$-locally testable linear property must have a $k$-characterization. In the case of affine-invariant linear families some special characterizations are known to lead to $k$-testability. We describe these special characterizations next.

**Definition 2.2** ($k$-single-orbit characterization). Let $C = (\overline{\alpha}, \{\overline{\lambda}_i\}_{i=1}^r)$ be a $k$-constraint on $\{\mathbb{F}_q^n \to \mathbb{F}_q\}$. The *orbit* of $C$ under the set of affine-transformations is the set of $k$-constraints $\{T \circ C\}_T = \left\{ ((T(\alpha_1), \ldots, T(\alpha_k)), \{\overline{\lambda}_i\}_{i=1}^r) \mid T : \mathbb{F}_q^n \to \mathbb{F}_q^n \text{ is an affine-transformation} \right\}$. We say that $C$ is a *$k$-single-orbit characterization* of $\mathcal{F}$ if the orbit of $C$ forms a $k$-characterization of $\mathcal{F}$.

The following theorem, due to Kaufman and Sudan [KS07], says that $k$-single-orbit characterization implies local testability.

**Theorem 2.3** (Single-orbit characterizability implies local testability, [KS07, Lemma 2.9]). *Let $\mathcal{F} \subseteq \{\mathbb{F}_q^n \to \mathbb{F}_q\}$ be an affine-invariant linear family. If $\mathcal{F}$ has a $k$-single-orbit characterization, then $\mathcal{F}$ has a $(k, \Omega(1/k^2))$-local tester.*

In view of the above theorem, it suffices to find a single-orbit characterization of $\mathrm{RM}[n, d, q]$ to test it. The following theorem, which we prove in the rest of this paper, thus immediately implies Theorem 1.1.

**Theorem 2.4.** *Let $q = p^s$ for prime $p$, and let $n, d$ be arbitrary positive integers. Then the Reed-Muller code $\mathrm{RM}[n, d, q]$ has a $k$-single-orbit characterization for $k \leq c_q \cdot \left(2^{p-1} + p - 1\right)^{(d+1)/(q(p-1))} \cdot q^{(d+1)/q}$ where $c_q \leq 3q^4$.*

## 2.2 Constraints vs. Monomials

One of the main simplifications derived from the study of affine-invariant linear properties is that it suffices to analyze the performance of constraints on "monomials" as opposed to general polynomials. This allows us to rephrase our target (a single-orbit characterization of $\mathrm{RM}[n, d, q]$) in somewhat simpler terms. Below we describe some of the essential notions, namely the "degree set", the "border set" and the relationship of these to single-orbit characterizations. This leads to a further reformulation of our main theorem as Theorem 2.13. Variations of most of the results and notions presented in this section appeared in previous works [KS07, GKS09, BS11, BGM+11]. In all the above works, with the exception of [KS07], the notions were specialized to the case of univariate funcions mapping $\mathbb{F}_{q^n}$ to $\mathbb{F}_q$ that are invariant over the set of affine-transformations over $\mathbb{F}_{q^n}$. In this work we focus on these notions in the context of affine-invariant linear properties over the domain $\mathbb{F}_q^n$.

---
[1] We note that as in [KS07] we do not require $A$ to be non-singular. Thus the affine-transformations we consider are not necessarily permutations from $\mathbb{F}_q^n$ to $\mathbb{F}_q^n$.

Let $\mathcal{F} \subseteq \{\mathbb{F}_q^n \to \mathbb{F}_q\}$ be a linear affine-invariant family of functions. Note that every member of $\{\mathbb{F}_q^n \to \mathbb{F}_q\}$ can be written uniquely as a polynomial in $\mathbb{F}_q[x_1, x_2, \ldots, x_n]$ of degree at most $q-1$ in each variable. For a monomial $\prod_{i=1}^n x_i^{d_i}$ over $n$ variables, we define its degree to be the vector $\overline{d} = (d_1, d_2, \ldots, d_n)$ and we define its *total degree* to be $\sum_{i=1}^n d_i$. For a function $f : \mathbb{F}_q^n \to \mathbb{F}_q$ we denote its *support*, denoted $\text{supp}(f)$, to be the set degrees in the support of the associated polynomial. I.e., $\text{supp}(f) = \{\overline{d} \in \{0, \ldots, q-1\}^n | c_{\overline{d}} \neq 0\}$ where $f(x) = \sum_{\overline{d}} c_{\overline{d}} x^{\overline{d}}$. The *degree set* $\text{Deg}(\mathcal{F})$ of $\mathcal{F}$ is simply the union of the supports of the functions in $\mathcal{F}$, i.e., $\text{Deg}(\mathcal{F}) = \cup_{f \in \mathcal{F}} \text{supp}(f)$.

While the degree set of the Reed-Muller codes are natural to study, they are also natural in more general contexts. The following lemma from [KS07] says that every affine-invariant linear property from $\mathbb{F}_q^n$ to $\mathbb{F}_q$ is uniquely determined by its degree set.

**Lemma 2.5** (Monomial extraction lemma, [KS07, Lemma 4.2]). *Let $\mathcal{F} \subseteq \{\mathbb{F}_q^n \to \mathbb{F}_q\}$ be an affine-invariant linear property. Then $\mathcal{F}$ has a monomial basis, that is, $\mathcal{F}$ is the set of all polynomials supported on monomials of the form $x^{\overline{d}}$ where $\overline{d} \in \text{Deg}(\mathcal{F})$.* [2]

One main structural feature of the degree sets of affine-invariant linear properties is that they are *p-shadow-closed*. Before giving the definition of a shadow-closed set of degrees we need to introduce a bit of notation. For a pair of integers $a, b$ let $a = \sum_j a_j p^j$, $b = \sum_j b_j p^j$ be their base-$p$ representation, respectively. We say that $b$ is in the *p-shadow* of $a$, and denote this $b \leq_p a$, if $b_j \leq a_j$ for all $j$. For a pair of integer vectors $\overline{d} = (d_1, d_2, \ldots, d_n)$, $\overline{e} = (e_1, e_2, \ldots, e_n)$ we say that $\overline{e} \leq_p \overline{d}$ if $e_i \leq d_i$ for every $i$.

**Definition 2.6** (Shadow-closed set of degrees). *For a vector of integers $\overline{d} = (d_1, d_2, \ldots, d_n)$ of length $n$, the p-shadow of $\overline{d}$ is the set $\text{Shadow}_p(\overline{d}) = \{\overline{e} = (e_1, e_2, \ldots, e_n) \mid \overline{e} \leq_p \overline{d}\}$. For a subset $S$ of integer vectors of length $n$ we let $\text{Shadow}_p(S) = \bigcup_{\overline{d} \in S} \text{Shadow}_p(\overline{d})$. Finally, we say that $S$ is p-Shadow-closed if $\text{Shadow}_p(S) = S$.*

The following lemma from [KS07] says that the degree set of every affine-invariant linear property over $\mathbb{F}_q^n$ is $p$-shadow-closed.

**Lemma 2.7** (Monomial spread lemma, [KS07, Lemma 4.6]). *let $\mathcal{F} \subseteq \{\mathbb{F}_{q^n} \to \mathbb{F}_q\}$ be an affine-invariant linear property. Then $\text{Deg}(\mathcal{F})$ is p-shadow-closed.*

A central element used in the proof of the above lemma and other aspects in the study of affine-invariant linear properties is Lucas's theorem, which we also need.

**Theorem 2.8** (Lucas's Theorem). *The monomial coefficient $\binom{n}{i}$ is non-zero mod $p$ if and only if $i \leq_p n$.*

---
[2] Our language is somewhat different from that of [KS07]. After translation, their lemma says that all monomials $x^{\overline{d}}$ are contained in $\mathcal{F}$. The other direction saying $\mathcal{F}$ is contained in the span of such monomials is immediate from the definition of $\text{Deg}(\mathcal{F})$.

The fact that the degree set of a linear affine-invariant family is $p$-shadow-closed motivates the notion of a "border" set, the set of minimal elements (under $\leq_p$) that are not in $\text{Deg}(\mathcal{F})$.

**Definition 2.9** (Border)**.** For an affine-invariant linear family $\mathcal{F} \subseteq \{\mathbb{F}_q^n \to \mathbb{F}_q\}$, its *border set*, denoted $\text{Border}(\mathcal{F})$, is the set

$$Border(\mathcal{F}) = \{\overline{e} \in \{0, \ldots, q-1\}^n | \overline{e} \notin \text{Deg}(\mathcal{F}) \text{ but } \forall \overline{e}' \leq_p \overline{e}, \overline{e}' \neq \overline{e}, \overline{e}' \in \text{Deg}(\mathcal{F})\}.$$

The relationship between the degree set and the border set of an affine-invariant linear family and single-orbit characterizability is given by the following lemma. This lemma says that for an affine-invariant linear family, in order to establish $k$-single-orbit characterizability it suffices to exhibit a $k$-constraint whose orbit accepts all monomials of the form $x^{\overline{d}}$ for $\overline{d} \in \text{Deg}(\mathcal{F})$ and rejects all monomials of the form $x^{\overline{b}}$ for $\overline{b} \in \text{Border}(\mathcal{F})$. It is similar in spirit to Lemma 3.2 of [BGM+11] which shows that a similar result holds for affine-invariant linear properties over $\mathbb{F}_{q^n}$.

**Lemma 2.10.** *Let $\mathcal{F} \subseteq \{\mathbb{F}_q^n \to \mathbb{F}_q\}$ be an affine-invariant linear property and let $C$ be a constraint. Then $C$ is a single-orbit characterization of $\mathcal{F}$ if the orbit of $C$ accepts every monomial $x^{\overline{d}}$ for $\overline{d} \in \text{Deg}(\mathcal{F})$ and rejects every monomial $x^{\overline{b}}$ for $\overline{b} \in \text{Border}(\mathcal{F})$.*

*Proof.* We need to show that for every affine-transformation $T : \mathbb{F}_q^n \to \mathbb{F}_q^n$ the constraint $T \circ C$ accepts all functions $f \in \mathcal{F}$, while for every $f \notin \mathcal{F}$ there exists an affine-transformation $T$ such that $T \circ C$ rejects $f$.

Since the set of monomials $x^{\overline{d}}$ for $\overline{d} \in \text{Deg}(\mathcal{F})$ forms a basis for $\mathcal{F}$, clearly we have that $C$ accepts all functions $f \in \mathcal{F}$. The fact that $\mathcal{F}$ is affine-invariant implies in turn that for every affine-transformation $T$ the constraint $T \circ C$ also accepts all functions $f \in \mathcal{F}$.

It remains to show that for every $f \notin \mathcal{F}$ there exists an affine-transformation $T : \mathbb{F}_q^n \to \mathbb{F}_q^n$ such that $T \circ C$ rejects $f$. Suppose in contrary that there exists a function $f \notin \mathcal{F}$ such that the orbit of $C$ accepts $f$, and let $\tilde{\mathcal{F}} = \{f : \mathbb{F}_q^n \to \mathbb{F}_q | T \circ C \text{ accepts } f \text{ for every affine-transformation } T\}$. Note that $\tilde{\mathcal{F}}$ is a linear affine-invariant property, and that our assumption on $f$ implies that $f \in \tilde{\mathcal{F}}$.

Since $f \notin \mathcal{F}$, and since $\text{Deg}(\mathcal{F})$ forms a basis for $\mathcal{F}$, there exists a monomial $x^{\overline{e}}$ in the support of $f$ such that $\overline{e} \notin \text{Deg}(\mathcal{F})$. The monomial extraction lemma (Lemma 2.5) then implies that $x^{\overline{e}}$ is also contained in $\tilde{\mathcal{F}}$. Let $\overline{b}$ be a minimal degree (with respect to $\leq_p$) such that $\overline{b} \leq_p \overline{e}$ and $\overline{b} \notin \text{Deg}(\mathcal{F})$. Then from the definition of the border we have that $\overline{b} \in \text{Border}(\mathcal{F})$. Furthermore, since $\tilde{\mathcal{F}}$ is linear and affine-invariant and $\overline{e} \in \text{Deg}(\tilde{\mathcal{F}})$, the monomial spread lemma (Lemma 2.7) implies that $\overline{b} \in \text{Deg}(\tilde{\mathcal{F}})$ and in particular $x^{\overline{b}} \in \tilde{\mathcal{F}}$. But this implies in turn that $x^{\overline{b}}$ is accepted by the orbit of $C$, a contradiction to our assumption that all degrees in the Border of $\mathcal{F}$ are rejected by the orbit of $C$. $\square$

In order to describe the border of the Reed-Muller family we shall use the following definition.

**Definition 2.11.** For integer $d$, let $d_0, d_1, \ldots$, be its expansion in base-$p$, i.e., $d_j$'s satisfy $0 \leq d_j < p$ and $d = \sum_{j=0}^{\infty} d_j p^j$. Let $b_i(d) = p^i + \sum_{j=i}^{\infty} d_j p^j$.

Note that $b_i(d) > d$ for every $i$ and conversely, for every integer $e > d$ there exists an $i$ such that $b_i(d) \leq_p e$. The $b_i(d)$'s are useful in describing the border monomials of the Reed-Muller family, as formalized below.

**Proposition 2.12.** *For every $n, d, q$, where $q = p^s$ for a prime $p$, we have*

$$\text{Deg}(\text{RM}[n,d,q]) = \left\{ \overline{d} = (d_1, \ldots, d_n) \in \{0, \ldots, q-1\}^n \mid \sum_{j=1}^n d_j \leq d \right\} \text{ and}$$

$$\text{Border}(\text{RM}[n,d,q]) \subseteq \left\{ \overline{e} = (e_1, \ldots, e_n) \in \{0, \ldots, q-1\}^n \mid \sum_{j=1}^n e_j = b_i(d) \text{ for some } 0 \leq i \leq s \right\}.$$

*Proof.* The fact that the degree set contains all $\overline{d}$ with $\sum_{j=1}^n d_j \leq d$ is immediate from the definitions. Now consider $\overline{f}$ such that $\sum_{j=1}^n f_j > d$. To verify the correctness of the border, we wish to show that there exists $\overline{e} \leq_p \overline{f}$ and $0 \leq i \leq s$ such that $\sum_{j=1}^n e_j = b_i(d)$. Let $\ell$ be the least index such that $\sum_{j=1}^\ell f_j = f > d$. Note that $f \leq d + q - 1$, since $f_\ell \leq q - 1$. Now let $f^{(0)}, f^{(1)}, \ldots$ denote the base-$p$ expansion of $f$ and let $d^{(0)}, d^{(1)}, \ldots$ denote the base-$p$ expansion of $d$. Since $f > d$, there must exist a largest index $i$ such that $f^{(i)} > d^{(i)}$ and for all $j > i$, $f^{(j)} = d^{(j)}$. For this choice of $i$, note that $b_i(d) \leq_p f$ and one can reduce $f_j$'s to $e_j$'s so that $\overline{e} \leq_p \overline{f}$ and $\sum_{j=1}^n e_j = b_i(d)$.

It remains to be shown that $i \leq s$. For this part note that if $b_i(d) > b_{i-1}(d)$ then $b_i(d) \geq b_{i-1}(d) + p^{i-1}$. Thus for all $i > s$, we have either $b_i(d) = b_s(d)$ or $b_i(d) \geq b_s(d) + p^s > d + q$. But since $f \leq d + q - 1$ it follows that we never need to use $i > s$. $\square$

Combining Lemma 2.10 and Proposition 2.12 we have that Theorem 2.4 follows immediately from Theorem 2.13 below.

**Theorem 2.13.** *Let $q = p^s$ for a prime $p$. Then there exists a $k$-constraint $C$ whose orbit accepts all monomials of total degree at most $d$ and rejects all monomials of total degree $b_i(d)$ for $0 \leq i \leq s$, for $k \leq 3q^4 \cdot \left(2^{p-1} + p - 1\right)^{(d+1)/(q(p-1))} \cdot q^{(d+1)/q}$.*

The rest of this paper will be devoted to proving Theorem 2.13.

## 3 Canonical monomials and a new constraint

In this section we introduce the notion of "canonical monomials" of a given degree — very simplified monomials that appear in every affine-invariant linear property containing monomials of a given degree. We then give a constraint that rejects canonical monomials of some special degrees, while accepting all monomials of lower degrees. Later, in Section 4, we show how to use this to build a constraint whose orbit accepts all monomials of total degree at most $d$ while rejecting all monomials of total degree $b_i(d)$, which suffices to get Theorem 2.13.

**Definition 3.1** (Canonical monomials). *Let $q = p^s$ for a prime $p$. The canonical monomial of (total) degree $d$ over $\mathbb{F}_q$ is the monomial $\prod_{i=1}^\ell x_i^{d_i}$ which satisfies $\sum_{i=1}^\ell d_i = d$, $d_i = q - q/p$ for all $2 \leq i \leq \ell$, $0 \leq d_1 \leq q - 1$ and $d_1 + q - q/p > q - 1$.*

We note that [HSS11] used a different canonical monomial (cf. Definition 4.1., [HSS11]) for the construction of their improved tester for the Reed-Muller codes. Our different choice of canonical monomial is needed to construct single-orbit characterizations which improve on those given in [HSS11] in terms of the number of queries. The main property of the canonical monomial, that we will use in Section 4.3 to prove Theorem 2.13 is that every affine-invariant linear family that contains any monomial of total degree $d$ also contains the canonical monomial of degree $d$. (see Lemma 4.8). This will imply in turn that if we can find constraints that *reject* this canonical monomial their orbit will reject every monomial of total degree $d$.

## 3.1 A new constraint on monomials of total degree $< p(q - q/p)$

The main technical novelty in our paper is a $k$-constraint $C$ that accepts all monomials of total degree strictly less than $p(q - q/p)$ in $p$ variables but rejects the canonical monomial of degree $p(q - q/p)$ (note that the latter monomial also has $p$ variables) for $k = (2^{p-1} + p - 1)q^{p-1}$. We state the lemma below and devote the rest of this section to proving this lemma.

**Lemma 3.2** (Main technical lemma). *For every $q$ which is a power of a prime $p$ there exists a $k$-constraint $C$ which accepts all monomials of total degree smaller than $p(q - q/p)$ in $p$ variables and rejects the canonical monomial (in $p$ variables) of degree $p(q - q/p)$ over $\mathbb{F}_q$, where $k = (2^{p-1} + p - 1)q^{p-1}$.*

It will be convenient for us to represent the constraint $C$ as a $p$-variate polynomial over $\mathbb{F}_q$. More precisely, suppose that $P(x)$ is a $p$-variate polynomial $P(x) \in \mathbb{F}_q[x_1, x_2, \ldots, x_p]$ that is non-zero on at most $k$ points in $\mathbb{F}_q^p$. We associate with $P(x)$ the $k$-constraint $C = (\overline{\alpha}, \overline{\lambda})$, $\overline{\alpha} = (\alpha_1, \ldots, \alpha_k) \in (\mathbb{F}_q^p)^k$, $\overline{\lambda} = (\lambda_1, \ldots, \lambda_k) \in \mathbb{F}_q^k$, where the vector $\overline{\alpha}$ consists of all points in $\mathbb{F}_q^p$ on which $P(x)$ is non-zero and $\lambda_j = P(\alpha_j)$ for all $1 \leq j \leq k$. Clearly, for every function $f: \mathbb{F}_q^p \to \mathbb{F}_q$ it holds that

$$\sum_{j=1}^{k} \lambda_j f(\alpha_j) = \sum_{\beta_1, \ldots, \beta_p \in \mathbb{F}_q} P(\beta_1, \ldots, \beta_p) \cdot f(\beta_1, \ldots, \beta_p) \quad (1)$$

Thus we reduce the task of finding a $k$-constraint which accepts all monomials of total degree smaller than $p(q - q/p)$ and rejects the canonical monomial of degree $p(q - q/p)$ to the task of finding a $p$-variate polynomial $P(x) \in \mathbb{F}_q[x_1, x_2, \ldots, x_p]$ with at most $k$ non-zero points in $\mathbb{F}_q^p$ such that $\sum_{\beta_1, \ldots, \beta_p \in \mathbb{F}_q} P(\beta_1, \ldots, \beta_p) \cdot M(\beta_1, \ldots, \beta_p) = 0$ for every monomial in $p$ variables of total degree smaller than $p(q - q/p)$ and $\sum_{\beta_1, \ldots, \beta_p \in \mathbb{F}_q} P(\beta_1, \ldots, \beta_p) \cdot M(\beta_1, \ldots, \beta_p) \neq 0$ when $M(x)$ is the canonical monomial of degree $p(q - q/p)$.

We start by describing the polynomial $P(x)$. The best way to describe this polynomial is via the notion of *directional derivatives*. Let $f: \mathbb{F}_q \to \mathbb{F}_q$ be a function. Define the derivative of $f$ in direction $y \in \mathbb{F}_q$ as $f_y(x) = f(x + y) - f(x)$. Define the iterated derivatives as

$$f_{y_1, \ldots, y_d}(x) = (f_{y_1, \ldots, y_{d-1}})_{y_d}(x) = \sum_{I \subseteq [d]} (-1)^{|I|+1} f\left(x + \sum_{i \in I} y_i\right).$$

Let $f(x)$ be the polynomial $f(x) = x_p^{q-1}$. Our polynomial $P(x)$ will be defined as follows.

$$P(x) = \frac{f_{x_1, \ldots, x_{p-1}}(x_p)}{x_1 \cdots x_{p-1}} = \frac{\sum_{I \subseteq [p-1]} (-1)^{|I|+1} (x_p + \sum_{i \in I} x_i)^{q-1}}{x_1 \cdots x_{p-1}}. \quad (2)$$

To see that $P(X)$ is indeed a polynomial we need to show that $f_{x_1, \ldots, x_{p-1}}(x_p)$ is divisible by $x_1 \cdots x_{p-1}$. This follows from the following lemma.

**Lemma 3.3.** *Let $f$ be a univariate polynomial over $\mathbb{F}_q$. Then the polynomial $f_y(x)$ is divisible by $y$.*

*Proof.* Write $f(x) = \sum_d c_d x^d$. Then

$$f_y(x) = f(x + y) - f(x) = \sum_d c_d (x + y)^d - \sum_d c_d x^d = \sum_d c_d \sum_{i=0}^{d} \binom{d}{i} x^{d-i} y^i - \sum_d c_d x^d$$

$$= \sum_{d > 0} c_d \sum_{i=1}^{d} \binom{d}{i} x^{d-i} y^i.$$

Thus all monomials in $f_y(x)$ contain the variable $y$ and hence $f_y(x)$ is divisible by $y$. □

In order to prove our main technical Lemma 3.2 it suffices to show that the number of non-zero points of $P(x)$ in $\mathbb{F}_q^p$ is at most $(2^{p-1} + p - 1)q^{p-1}$, that it accepts all monomials in $p$ variables of total degree smaller $p(q - q/p)$, and that it rejects the canonical monomial of degree $p(q - q/p)$. We prove these three claims in Lemmas 3.4, 3.6 and 3.8 below, respectively. Given these three lemmas our main technical Lemma 3.2 is immediate. We start with bounding the number of non-zeros of $P(x)$.

**Lemma 3.4.** *The number of non-zero points of $P(x)$ in $\mathbb{F}_q^p$ is at most $(2^{p-1} + p - 1)q^{p-1}$.*

*Proof.* Let $\overline{\beta} = (\beta_1, \ldots, \beta_p) \in \mathbb{F}_q^p$. Suppose first that all first $p - 1$ coordinates of $\overline{\beta}$ are non-zero, so that the denominator of $P(\overline{\beta})$ is non-zero. Suppose furthermore that all terms of the form $(\beta_p + \sum_{i \in I} \beta_i)^{q-1}$ in the numerator of $P(\overline{\beta})$ are non-zero. Then in this case we have that

$$P(\overline{\beta}) = \frac{\sum_{I \subseteq [p-1]}(-1)^{|I|+1} \cdot 1}{\beta_1 \cdots \beta_{p-1}} = -\frac{\sum_{i=0}^{p-1}\binom{p-1}{i}(-1)^i}{\beta_1 \cdots \beta_{p-1}} = 0.$$

Thus we have that whenever $\beta_1, \ldots, \beta_{p-1}$ are all non-zero, $P(\overline{\beta})$ can be non-zero only if at least one of the terms of the form $(\beta_p + \sum_{i \in I} \beta_i)^{q-1}$ in its numerator equals zero. Note that each such term has exactly $q^{p-1}$ assignments in $\mathbb{F}_q^p$ that make it zero. Since the number of terms in the numerator is $2^{p-1}$, the total number of points in $\mathbb{F}_q^p$ that satisfy that at least one of the terms in the numerator equals zero is at most $2^{p-1} \cdot q^{p-1}$.

Concluding, we have that there are at most $2^{p-1}q^{p-1}$ vectors $\overline{\beta} \in \mathbb{F}_q^p$ which satisfy that $\beta_1, \beta_2, \ldots, \beta_{p-1}$ are all non-zero and in addition $P(\overline{\beta}) \neq 0$. Since there are at most $(p-1)q^{p-1}$ elements $\overline{\beta} \in \mathbb{F}_q^p$ in which at least one of the first $p - 1$ coordinates is zero, we conclude that the number of elements $\overline{\beta} \in \mathbb{F}_q^p$ such that $P(\overline{\beta}) \neq 0$ is at most $2^{p-1}q^{p-1} + (p-1)q^{p-1} = (2^{p-1} + p - 1)q^{p-1}$. □

Next we show that the constraint $C$ associated with $P(x)$ accepts all monomials in $p$ variables of total degree smaller than $p(q - q/p)$. For this we need the following well-known fact.

**Fact 3.5.** *Let $q$ be a prime-power, and let $i$ be an integer in $\{0, 1, \ldots, q - 1\}$. Then*

$$\sum_{\beta \in \mathbb{F}_q} \beta^i = \begin{cases} -1 & i = q - 1 \\ 0 & \text{otherwise} \end{cases}$$

*Proof.* Recall that the multiplicative group of $\mathbb{F}_q^*$ is cyclic and let $\gamma$ be a generator of this group. Then for all $i \in \{0, 1, \ldots, q - 2\}$ we have

$$\sum_{\beta \in \mathbb{F}_q} \beta^i = \sum_{j=1}^{q-1} \gamma^j = \frac{\gamma^{q-1} - 1}{\gamma - 1} = \frac{1 - 1}{\gamma - 1} = 0,$$

whereas for $i = q - 1$ we have that

$$\sum_{\beta \in \mathbb{F}_q} \beta^{q-1} = \sum_{\beta \in \mathbb{F}_q^*} 1 = q - 1 = -1.$$

□

**Lemma 3.6.** *Let $C$ be the constraint associated with $P(x)$. Then $C$ accepts all monomials in $p$ variables of total degree smaller than $p(q - q/p)$.*

*Proof.* Let $m$ be a monomial in $p$ variables of total degree $d < p(q - q/p)$. We shall show that $\sum_{\beta_1,\ldots,\beta_p \in \mathbb{F}_q} m(\beta_1, \beta_2, \ldots, \beta_p) \cdot m'(\beta_1, \beta_2, \ldots, \beta_p) = 0$ for every monomial $m'$ in $P(x)$. This will show in turn that $\sum_{\beta_1,\ldots,\beta_p \in \mathbb{F}_q} m(\beta_1,\ldots,\beta_p) \cdot P(\beta_1,\ldots,\beta_p) = 0$ and hence $C$ accepts $m$.

Let $m'$ be a monomial in $P(x)$. First note that all monomials in the numerator of $P(x)$ have total degree exactly $q-1$ and hence all monomials in $P(x)$ have total degree $q-1-(p-1) = q-p$. Thus $m'$ has total degree $q-p$, and $m \cdot m'$ is a monomial of total degree $q-p+d < q-p+p(q-q/p) = p(q-1)$. Since $m \cdot m'$ is a monomial in $p$ variables, by pigeonhole principle there exists a variable $x_i$ in $m \cdot m'$ of degree smaller than $q-1$. Without loss of generality suppose that $x_1$ has degree $d' < q-1$ in $m \cdot m'$, and let $m \cdot m' = x_1^{d'} \cdot \prod_{i=2}^{p} x_i^{d_i}$.

Thus we have

$$\sum_{\beta_1,\ldots,\beta_p \in \mathbb{F}_q} (m \cdot m')(\beta_1,\ldots,\beta_p) = \sum_{\beta_1,\ldots,\beta_p \in \mathbb{F}_q} \beta_1^{d'} \cdot \prod_{i=2}^{p} \beta_i^{d_i} = \left( \sum_{\beta_1 \in \mathbb{F}_q} \beta_1^{d'} \right) \prod_{i=2}^{p} \left( \sum_{\beta_i \in \mathbb{F}_q} \beta_i^{d_i} \right) = 0,$$

where the last equality follows from Fact 3.5 above and the fact that $d' < q - 1$. □

We complete the proof of Lemma 3.2 by proving that the constraint $C$ associated with $P(x)$ rejects the canonical monomial of degree $p(q - q/p)$. In order to prove this theorem we shall use Kummer's Theorem which generalizes Lucas's Theorem (Theorem 2.8) and gives a condition for when a multinomial coefficient $\binom{n}{\gamma_1, \gamma_2, \ldots, \gamma_k} = \frac{n!}{\gamma_1! \gamma_2! \cdots \gamma_k!} \not\equiv 0 \mod p$.

**Theorem 3.7** (Kummer's Theorem, [Kum36]). *Let $n, \gamma_1, \gamma_2, \ldots, \gamma_k$ be integers such that $n = \gamma_1 + \gamma_2 + \ldots + \gamma_k$. Then the multinomial coefficient $\binom{n}{\gamma_1, \gamma_2, \ldots, \gamma_k} = \frac{n!}{\gamma_1! \gamma_2! \cdots \gamma_k!} \not\equiv 0 \mod p$ if and only if $\gamma_1, \gamma_2, \ldots, \gamma_k$ sum to $n$ in base-$p$ without carry.*

**Lemma 3.8.** *Let $C$ be the constraint associated with $P(x)$. Then $C$ rejects the canonical monomial of degree $p(q - q/p)$ over $\mathbb{F}_q$.*

*Proof.* The canonical monomial of degree $p(q - q/p)$ over $\mathbb{F}_q$ is the monomial $m = \prod_{i=1}^{p} x_i^{q-q/p}$. Let $m' = \prod_{i=1}^{p} x_i^{q/p-1}$. We claim that $\sum_{\beta_1,\ldots,\beta_p \in \mathbb{F}_q} (m \cdot m')(\beta_1,\ldots,\beta_p) \neq 0$ while for every other monomial $m'' \neq m'$ in the support of $P(x)$ it holds that $\sum_{\beta_1,\ldots,\beta_p \in \mathbb{F}_q} (m \cdot m'')(\beta_1,\ldots,\beta_p) = 0$. Thus in order to prove the lemma it will suffice to show that the monomial $m'$ is in the support of $P(x)$.

We start by showing that $\sum_{\beta_1,\ldots,\beta_p \in \mathbb{F}_q} (m \cdot m')(\beta_1,\ldots,\beta_p) \neq 0$.

$$\sum_{\beta_1,\ldots,\beta_p \in \mathbb{F}_q} (m \cdot m')(\beta_1,\ldots,\beta_p) = \sum_{\beta_1,\ldots,\beta_p \in \mathbb{F}_q} \prod_{i=1}^{p} \beta_i^{q-1} = \prod_{i=1}^{p} \left( \sum_{\beta_i \in \mathbb{F}_q} \beta_i^{q-1} \right) = (-1)^p \neq 0,$$

where the last equality follows from Fact 3.5 above.

Next we show that for every monomial $m'' \neq m'$ in the support of $P(x)$ it holds that $\sum_{\beta_1,\ldots,\beta_p \in \mathbb{F}_q} (m \cdot m'')(\beta_1,\ldots,\beta_p) = 0$. Let $m'' \neq m'$ be a monomial in the support of $P(x)$. Then $m''$ is a monomial of total degree $q - p$ and the fact that $m'' \neq m'$ implies that $m''$ has a variable of degree smaller than $(q-p)/p = q/p - 1$. Without loss of generality suppose that the variable $x_1$ has degree smaller than $q/p - 1$ in $m''$ and note that this implies that the variable $x_1$ has degree $d' < q - 1$ in $m \cdot m''$. Let $m \cdot m'' = x_1^{d'} \cdot \prod_{i=2}^{p} x_i^{d_i}$. Then we have

$$\sum_{\beta_1,\ldots,\beta_p \in \mathbb{F}_q} (m \cdot m'')(\beta_1,\ldots,\beta_p) = \sum_{\beta_1,\ldots,\beta_p \in \mathbb{F}_q} \beta_1^{d'} \prod_{i=2}^{p} \beta_i^{d_i} = \left( \sum_{\beta_1 \in \mathbb{F}_q} \beta_1^{d'} \right) \prod_{i=2}^{p} \left( \sum_{\beta_i \in \mathbb{F}_q} \beta_i^{q-1} \right) = 0,$$

where the last equality holds since $\sum_{\beta_1 \in \mathbb{F}_q} \beta_1^{d'} = 0$ due to Fact 3.5 above and the fact that $d' < q-1$.

We have shown that $\sum_{\beta_1,\ldots,\beta_p \in \mathbb{F}_q}(m \cdot m')(\beta_1,\ldots,\beta_p) \neq 0$ while for every other monomial $m'' \neq m'$ in the support of $P(x)$ it holds that $\sum_{\beta_1,\ldots,\beta_p \in \mathbb{F}_q}(m \cdot m'')(\beta_1,\ldots,\beta_p) = 0$, and hence in order to prove the lemma it suffices to show that the monomial $m'$ is in the support of $P(x)$. This happens in turn if and only if the monomial $\tilde{m} = x_1^{q/p-1} \prod_{i=2}^{p} x_i^{q/p}$ is in the numerator of $P(x)$. Note that of all terms of the form $(x_p + \sum_{i \in I} x_i)^{q-1}$ in the numerator of $P(x)$, the monomial $\tilde{m}$ can only belong to the support of $(x_p + \sum_{i \in [p-1]} x_i)^{q-1} = (x_1 + x_2 + \ldots + x_p)^{q-1}$. Thus it suffices to show that the monomial $\tilde{m}$ belongs to the support of $(x_1 + x_2 + \ldots + x_p)^{q-1}$.

In order to show the above we resort to Kummer's Theorem. Expanding $(x_1 + x_2 + \ldots + x_p)^{q-1}$ we have that the coefficient of the monomial $\tilde{m}$ is $\binom{q-1}{\gamma_1,\ldots,\gamma_p}$ for $\gamma_1 = q/p - 1$ and $\gamma_i = q/p$ for all $2 \leq i \leq p$. Noting that $\gamma_1,\ldots,\gamma_p$ sum to $q-1$ without carry in base-$p$, Kummer's Theorem implies that $\binom{q-1}{\gamma_1,\ldots,\gamma_p}$ is non-zero mod $p$. This implies in turn that $\tilde{m}$ is contained in the support of the polynomial $(x_1 + x_2 + \ldots + x_p)^{q-1}$, thus completing the proof of the lemma. □

Given Lemmas 3.4, 3.6 and 3.8 the proof of Lemma 3.2 is immediate.

*Proof of Lemma 3.2.* Let $P(x)$ be the polynomial given in (2), and let $C$ be the constraint on $\{\mathbb{F}_q^p \to \mathbb{F}_q\}$ associated with $P(x)$. From Lemma 3.4 we have that the number of non-zero points of $P(x)$ in $\mathbb{F}_q^p$ is at most $(2^{p-1} + p - 1)q^{p-1}$, and hence $C$ is a $((2^{p-1} + p - 1)q^{p-1})$-constraint. Lemma 3.6 implies that $C$ accepts all monomials of total degree smaller than $p(q - q/p)$, while Lemma 3.8 implies that $C$ rejects the canonical monomial of degree $p(q - q/p)$. □

## 3.2 Tightness of our analysis

Next we show that the upper bound on the number of non-zero points of $P(x)$ in $\mathbb{F}_q^p$ given in Lemma 3.4 is essentially tight.

**Lemma 3.9.** *The number of non-zero points of $P(x)$ in $\mathbb{F}_p^q$ is at least*

$$(2^{p-1} - p - 1)q^{p-1} - 2^{p-1}(2^{p-1} - 1)q^{p-2}.$$

*Proof.* We will show that the numerator of $P(x)$ is non-zero for at least $2^{p-1}q^{p-1} - 2^{p-1}(2^{p-1}-1)q^{p-2}$ elements in $\mathbb{F}_q^p$. Since the denominator of $P(x)$ is zero for at most $(p-1)q^{p-1}$ elements in $\mathbb{F}_q^p$ this will show that $P(x)$ is non-zero for at least $2^{p-1}q^{p-1} - 2^{p-1}(2^{p-1}-1)q^{p-2} - (p-1)q^{p-1}$ points in $\mathbb{F}_q^p$.

Let $\overline{\beta} = (\beta_1, \ldots, \beta_p)$ be a random point in $\mathbb{F}_q^p$. Let $E$ be the event that the numerator of $P(\overline{\beta})$ is non-zero. Our goal will be to show that $\Pr[E] \geq 2^{p-1}/q - 2^{p-1}(2^{p-1}-1)/q^2$. For a subset $I \subseteq [p-1]$ let $E_I$ be the event that $(\beta_p + \sum_{i \in I} \beta_i)^{q-1} = 1$. Let $E'$ be the event that exactly one of the events $E_I$ holds. Clearly, $E' \subseteq E$ and hence it suffices to show that $\Pr[E'] \geq 2^{p-1}/q - 2^{p-1}(2^{p-1}-1)/q^2$.

Note that $\Pr[E_I] = 1/q$ for all $I \subseteq [p-1]$ and $\Pr[E_I \cap E_J] = 1/q^2$ for all $I \neq J$, $I, J \subseteq [p-1]$ since $\beta_p + \sum_{i \in I} \beta_i = 0$ and $\beta_p + \sum_{j \in J} \beta_j = 0$ are linearly independent linear equations.

Thus have that

$$\begin{aligned}
\Pr[E'] &\geq \sum_{I \subseteq [p-1]} \left( \Pr[E_I] - \sum_{J \subseteq [p-1], J \neq I} \Pr[E_I \cap E_J] \right) \\
&= \sum_{I \subseteq [p-1]} \left( 1/q - \sum_{J \subseteq [p-1], J \neq I} 1/q^2 \right) \\
&= 2^{p-1}(1/q - (2^{p-1} - 1)/q^2) \\
&= 2^{p-1}/q - 2^{p-1}(2^{p-1} - 1)/q^2
\end{aligned}$$

□

## 4 Proof of Theorem 2.13

In this section we use Lemma 3.2 to prove Theorem 2.13. This part is done in several steps. In Section 4.1 we extend the constraint from the previous section to get constraints rejecting canonical monomials of degree $d+1$, while accepting all monomials of total degree at most $d$ for an arbitrary integer $d$. Next, in Section 4.2 we combine the various constraints from the previous step to find one constraint which rejects all the canonical monomials of degree $b_i(d)$ for every $d$, while accepting all monomials of total degree at most $d$. Finally, in Section 4.3, we prove Theorem 2.13. In this part, we use some of the standard facts about affine-invariance to conclude that the orbit of the constraint from the previous step must reject all monomials $x^{\overline{b}}$ for $\overline{b} \in \text{Border}(\text{RM}[n,d,q])$ while accepting all monomials $x^{\overline{d}}$ for $\overline{d} \in \text{Deg}(\text{RM}[n,d,q])$.

### 4.1 Rejecting canonical monomials of arbitrary degree $d+1$

We start by showing how the constraint guaranteed by Lemma 3.2 can be turned, via an operation on constraints that we call the *convolution operation*, into a constraint which rejects the canonical monomial of degree $d+1$ and accepts all monomials of total degree at most $d$ for an arbitrary integer $d$. This step is given in the following lemma.

**Lemma 4.1.** *For every $q$ which is a power of a prime $p$, and for every integers $d$, $n$ there exists a $k$-constraint $C$ on $\{\mathbb{F}_q^n \to \mathbb{F}_q\}$ which accepts all monomials of total degree at most $d$ and rejects the canonical monomial of degree $d+1$ over $\mathbb{F}_q$ for $k \leq q^2 \cdot \left(2^{p-1} + p - 1\right)^{(d+1)/(q(p-1))} \cdot q^{(d+1)/q}$.*

For the proof of the above lemma first introduce the convolution operation. For a pair of vectors $\gamma = (\gamma_1, \ldots, \gamma_{n_1})$ and $\gamma' = (\gamma'_1, \ldots, \gamma'_{n_2})$ let $\gamma \circ \gamma' = (\gamma_1, \ldots, \gamma_{n_1}, \gamma'_1, \ldots, \gamma'_{n_2})$ denote their concatenation.

**Definition 4.2** (Convolution of constraints). Let $C_1 = \left(\overline{\alpha}^{(1)}, \{\overline{\lambda}_i^{(1)}\}_{i=1}^{r_1}\right)$ be a $k_1$-constraint on $\{\mathbb{F}_q^{n_1} \to \mathbb{F}_q\}$, and let $C_2 = \left(\overline{\alpha}^{(2)}, \{\overline{\lambda}_i^{(2)}\}_{i=1}^{r_2}\right)$ be a $k_2$-constraint on $\{\mathbb{F}_q^{n_2} \to \mathbb{F}_q\}$. Their convolution $C = C_1 \otimes C_2$ is the $(k_1 \cdot k_2)$-constraint $C = \left(\overline{\alpha}, \{\overline{\lambda}_{(i_1,i_2)}\}_{i_1=1,i_2=1}^{r_1,r_2}\right)$ on $\{\mathbb{F}_q^{n_1+n_2} \to \mathbb{F}_q\}$, where $\overline{\alpha} \in (\mathbb{F}_q^{n_1+n_2})^{k_1 \times k_2}$ and $\overline{\lambda}_{(i_1,i_2)} \in (\mathbb{F}_q)^{k_1 \times k_2}$ for all $1 \leq i_1 \leq r_1$, $1 \leq i_2 \leq r_2$, and are defined as follows:

$$\alpha_{(j_1,j_2)} = \alpha_{j_1}^{(1)} \circ \alpha_{j_2}^{(2)}, \quad \lambda_{(i_1,i_2),(j_1,j_2)} = \lambda_{i_1,j_1}^{(1)} \cdot \lambda_{i_2,j_2}^{(2)}$$

for all $1 \leq j_1 \leq k_1$, $1 \leq j_2 \leq k_2$, $1 \leq i_1 \leq r_1$, $1 \leq i_2 \leq r_2$.

**Proposition 4.3.** *Let $C_1$ be a $k_1$-constraint on $\{\mathbb{F}_q^{n_1} \to \mathbb{F}_q\}$ which accepts all monomials $x^{\overline{d}}$ with $\overline{d} \in D_1$ and rejects all monomials $x^{\overline{b}}$ with $\overline{b} \in B_1$, and let $C_2$ be a $k_2$-constraint on $\{\mathbb{F}_q^{n_2} \to \mathbb{F}_q\}$ which accepts all monomials $x^{\overline{d}}$ with $\overline{d} \in D_2$ and rejects all monomials $x^{\overline{b}}$ with $\overline{b} \in B_2$. Then $C_1 \otimes C_2$ is a $(k_1 \cdot k_2)$-constraint on $\{\mathbb{F}_q^{n_1+n_2} \to \mathbb{F}_q\}$ which accepts all monomials $x^{\overline{d}_1 \circ \overline{e}_2}$ and $x^{\overline{e}_1 \circ \overline{d}_2}$ where $\overline{d}_1 \in D_1$, $\overline{d}_2 \in D_2$ and $\overline{e}_1 \in \{0,\ldots,q-1\}^{n_1}$, $\overline{e}_2 \in \{0,\ldots,q-1\}^{n_2}$ are arbitrary, and rejects all monomials of the form $x^{\overline{b}_1 \circ \overline{b}_2}$ where $\overline{b}_1 \in B_1$ and $\overline{b}_2 \in B_2$.*

*Proof.* Let $\overline{e} = \overline{e}_1 \circ \overline{e}_2$ where $\overline{e}_1 \in \{0,\ldots,q-1\}^{n_1}$, $\overline{e}_2 \in \{0,\ldots,q-1\}^{n_2}$. Then for every $1 \leq i_1 \leq r_1$, $1 \leq i_2 \leq r_2$ we have that

$$\sum_{j_1=1}^{k_1} \sum_{j_2=1}^{k_2} \lambda_{(i_1,i_2),(j_1,j_2)} (\alpha_{(j_1,j_2)})^{\overline{e}} = \sum_{j_1=1}^{k_1} \sum_{j_2=1}^{k_2} \lambda_{i_1,j_1}^{(1)} \cdot \lambda_{i_2,j_2}^{(2)} \cdot \left(\alpha_{j_1}^{(1)} \circ \alpha_{j_2}^{(2)}\right)^{\overline{e}}$$

$$= \sum_{j_1=1}^{k_1} \sum_{j_2=1}^{k_2} \lambda_{i_1,j_1}^{(1)} \cdot \lambda_{i_2,j_2}^{(2)} \cdot \left(\alpha_{j_1}^{(1)}\right)^{\overline{e}_1} \left(\alpha_{j_2}^{(2)}\right)^{\overline{e}_2}$$

$$= \left( \sum_{j_1=1}^{k_1} \lambda_{i_1,j_1}^{(1)} \left(\alpha_{j_1}^{(1)}\right)^{\overline{e}_1} \right) \cdot \left( \sum_{j_2=1}^{k_2} \lambda_{i_2,j_2}^{(2)} \left(\alpha_{j_2}^{(2)}\right)^{\overline{e}_2} \right).$$

Hence $C$ accepts all monomials of the form $x^{\overline{d}_1 \circ \overline{e}_2}$ and $x^{\overline{e}_1 \circ \overline{d}_2}$ where $\overline{d}_1 \in D_1$, $\overline{d}_2 \in D_2$ and $\overline{e}_1 \in \{0,\ldots,q-1\}^{n_1}$, $\overline{e}_2 \in \{0,\ldots,q-1\}^{n_2}$ are arbitrary and rejects all monomials of the form $x^{\overline{b}_1 \circ \overline{b}_2}$ where $\overline{b}_1 \in B_1$, $\overline{b}_2 \in B_2$. □

The convolution operation, applied to the constraint given by Lemma 3.2, suffices for proving Lemma 4.1 when $p(q - q/p)$ divides $d+1$. However, since this is not always the case we need to consider also testing univariate monomials of degree at most $q-2$. The following lemma covers this case.

**Lemma 4.4** (Testing univariate monomials of degree at most $q-2$, (cf. [RS96])). *Let $d$ be an integer in $\{0,1,\ldots,q-2\}$. Then there exists a $(d+2)$-constraint $C$ on $\{\mathbb{F}_q \to \mathbb{F}_q\}$ which accepts all monomials $x^e$ for $e \leq d$ and rejects the monomial $x^{d+1}$.*

*Proof.* Let $C = (\overline{\alpha}, \overline{\lambda})$ be the $(d+2)$-constraint defined as follows. Let $\alpha_1, \ldots, \alpha_{d+2} \in \mathbb{F}_q$ be distinct elements (note that they do exist since $d+2 \leq q$). Let $\overline{\lambda} = (\lambda_1, \ldots, \lambda_{d+2}) \in \mathbb{F}_q^{d+2}$ be a non-zero vector satisfying $\sum_{i=1}^{d+2} \lambda_i \alpha_i^{\ell} = 0$ for all $\ell \in \{0,\ldots,d\}$. Note that such a vector $\overline{\lambda}$ exists since each of the constraints above is a homogenous linear constraint on $\overline{\lambda}$ and there are only $d+1$ such constraints and $d+2$ variables. We claim that $\sum_{i=1}^{d+2} \lambda_i \alpha_i^{d+1} \neq 0$, since if it were then $\overline{\lambda}$ would be in the null space of the Vandermonde matrix $[\alpha_i^j]_{i=1,j=0}^{d+2,d+1}$. Thus $C$ rejects $x^{d+1}$ while accepting $x^e$ for every $e \leq d$. □

Given Proposition 4.3 and Lemma 4.4 we are ready to prove Lemma 4.1.

*Proof of Lemma 4.1.* Write $d+1$ as $d+1 = r + \ell(q - q/p)$ where $0 \leq r \leq q-1$ and $r + q - q/p > q-1$. Write $\ell = \ell' p + r'$ where $0 \leq r' < p$. Let $C_1$ be the $(r+1)$-constraint guaranteed by Lemma 4.4 for the degree $r-1$, and let $C_2$ be the $(q - q/p + 1)$-constraint guaranteed by the same lemma for the degree $q - q/p - 1$. Let $C_3$ be the $k'$-constraint guaranteed by Lemma 3.2. Finally, let $C$ be the $((r+1) \cdot (q - q/p + 1)^{r'} \cdot (k')^{\ell'})$-constraint which is the convolution of $C_1$ with $r'$ copies of $C_2$ and $\ell'$ copies of $C_3$. That is, $C = C_1 \otimes C_2^{\otimes r'} \otimes C_3^{\ell'}$. We claim that $C$ accepts all monomials of total degree at most $d$ and rejects the canonical monomial of degree $d+1$ (if the arity of $C = \left(\overline{\alpha}, \{\overline{\lambda}_i\}_{i=1}^r\right)$ is smaller than $n$ then we extend $C$ to be of arity $n$ by concatenating sufficient number of 1's to each element $\alpha_j$ in the vector $\overline{\alpha}$).

To see this suppose first that $m = x_1^{d_1} \cdots x_n^{d_n}$ is a monomial of total degree at most $d$. In this case we have that either one of the variables $x_2, \ldots, x_{\ell+1}$ is of degree smaller than $q - q/p$ or that the variable $x_1$ is of degree smaller than $r$. From Proposition 4.3 this implies that the constraint $C$ accepts the monomial $m$. Suppose on the other hand that $m$ is the canonical monomial of degree $d+1$. Then in this case we have that all of the variables $x_2, \ldots, x_{\ell+1}$ are of degree $q - q/p$ and the variable $x_1$ is of degree $r$. Hence Proposition 4.3 implies that $C$ rejects the monomial $m$.

Finally note that the locality of $C$ is $k = (r+1) \cdot (q - q/p + 1)^{r'} \cdot (k')^{\ell'}$ where $k' = (2^{p-1} + p - 1)q^{p-1}$, $\ell' \leq (d+1)/((q-q/p)p)$, $r \leq q-1$ and $r' \leq p$. Using the following series of simplifications, we can bound $k$ as claimed.

$$\begin{aligned} k &= (r+1) \cdot (q - q/p + 1)^{r'} \cdot ((2^{p-1} + p - 1)q^{p-1})^{\ell'} \\ &\leq q \cdot (q)^{r'} \cdot ((2^{p-1} + p - 1)q^{p-1})^{\ell'} \\ &= q \cdot (2^{p-1} + p - 1)^{\ell'} \cdot q^{r' + (p-1) \cdot \ell'} \\ &\leq q \cdot (2^{p-1} + p - 1)^{\ell'} \cdot q^{((p-1)/p) \cdot \ell + 1} \\ &\leq q^2 \cdot (2^{p-1} + p - 1)^{(d+1)/((q-q/p)p)} \cdot q^{((p-1)/p) \cdot (d/(q-q/p))} \\ &= q^2 \cdot (2^{p-1} + p - 1)^{(d+1)/(q(p-1))} \cdot q^{(d+1)/q}. \end{aligned}$$

$\square$

## 4.2 Rejecting all canonical monomials in the border simultaneously

Next we show for every integer $d$ the existence of a $k$-constraint which accepts all monomials of total degree at most $d$ and rejects all the canonical monomials of degree $b_i(d)$ for $0 \leq i \leq s$ simultaneously. For proving this we shall use the union operation on constraints defined as follows. For an integer $k$, let $0_k$ denote the all-zeros vector of length $k$.

**Definition 4.5** (Union of constraints). Let $C_1 = \left(\overline{\alpha}^{(1)}, \{\overline{\lambda}_i^{(1)}\}_{i=1}^{r_1}\right)$ be a $k_1$-constraint on $\{\mathbb{F}_q^n \to \mathbb{F}_q\}$ and let $C_2 = \left(\overline{\alpha}^{(2)}, \{\overline{\lambda}_i^{(2)}\}_{i=1}^{r_2}\right)$ be a $k_2$-constraint on $\{\mathbb{F}_q^n \to \mathbb{F}_q\}$. Their union $C = C_1 \cup C_2$ is the $(k_1 + k_2)$-constraint $C = \left(\overline{\alpha}, \{\overline{\lambda}_i'^{(1)}\}_{i=1}^{r_1} \cup \{\overline{\lambda}_i'^{(2)}\}_{i=1}^{r_2}\right)$ on $\{\mathbb{F}_q^n \to \mathbb{F}_q\}$ defined by

$$\overline{\alpha} = \overline{\alpha}^{(1)} \circ \overline{\alpha}^{(2)},$$

$$\overline{\lambda}_i'^{(1)} = \overline{\lambda}_i^{(1)} \circ 0_{k_2} \text{ for all } 1 \leq i \leq r_1,$$

and

$$\overline{\lambda}_i'^{(2)} = 0_{k_1} \circ \overline{\lambda}_i^{(2)} \text{ for all } 1 \leq i \leq r_2.$$

**Proposition 4.6.** *Let $C_1$ be a constraint which accepts all monomials with degrees in $D_1$ and rejects all monomials with degrees in $B_1$, and let $C_2$ be a constraint which accepts all monomials with degrees in $D_2$ and rejects all monomials with degrees in $B_2$. Then $C_1 \cup C_2$ accepts all monomials with degrees in $D_1 \cap D_2$ and rejects all monomials with degrees in $B_1 \cup B_2$.*

*Proof.* For every degree $\overline{d}$,

$$\sum_{j=1}^{k_1+k_2} \lambda'^{(1)}_{i,j} \alpha_j^{\overline{d}} = \sum_{j=1}^{k_1} \lambda^{(1)}_{i,j} \left(\alpha_j^{(1)}\right)^{\overline{d}} \text{ for all } 1 \le i \le r_1,$$

and similarly

$$\sum_{j=1}^{k_1+k_2} \lambda'^{(2)}_{i,j} \alpha_j^{\overline{d}} = \sum_{j=1}^{k_2} \lambda^{(2)}_{i,j} \left(\alpha_j^{(2)}\right)^{\overline{d}} \text{ for all } 1 \le i \le r_2.$$

Thus the constraint $C$ accepts monomials of degree $\overline{d}$ if and only if both $C_1$ and $C_2$ accept the monomial of degree $\overline{d}$. Hence $C$ accepts all monomials with degrees in $D_1 \cap D_2$ and rejects all monomials with degrees in $B_1 \cup B_2$. □

Given the above proposition we can now build a constraint which accepts all monomials of total degree at most $d$ while rejecting all the canonical monomials of degree $b_i(d)$ for $0 \le i \le s$.

**Lemma 4.7.** *Let $q = p^s$ for a prime $p$ and let $n, d$ be arbitrary positive integers. Recall the definition of the integers $b_i(d)$ given in Definition 2.11. Then there exists a $k$-constraint $C$ on $\{\mathbb{F}_q^n \to \mathbb{F}_q\}$ which accepts all monomials of total degree at most $d$ and rejects all canonical monomials of degree $b_i(d)$ for $0 \le i \le s$, where $k \le 3q^4 \cdot (2^{p-1} + p - 1)^{(d+1)/(q(p-1))} \cdot q^{(d+1)/q}$.*

*Proof.* For all $0 \le i \le s$ let $C_i$ be the $k_i$-constraint which accepts all monomials of total degree at most $b_i(d) - 1$ and rejects the canonical monomial of degree $b_i(d)$ over $\mathbb{F}_q$ as guaranteed by Lemma 4.1 with $k_i \le q^2 \cdot (2^{p-1}+p-1)^{(d+1+q)/(q(p-1))} \cdot q^{(d+1+q)/q} \le 3q^3 \cdot (2^{p-1}+p-1)^{(d+1)/(q(p-1))} \cdot q^{(d+1)/q}$. Let $C = \bigcup_{i=0}^{s} C_i$. Then $C$ is a $k$-constraint for $k = \sum_{i=0}^{s} k_i \le 3q^4 \cdot (2^{p-1} + p - 1)^{(d+1)/(q(p-1))} \cdot q^{(d+1)/q}$. Proposition 4.6 implies that $C$ accepts all monomials of total degree at most $d$ and rejects all canonical monomials of degree $b_i(d)$ for $0 \le i \le s$, giving the claimed assertion. □

### 4.3 Rejecting all monomials in the border

In order to complete the prof of Theorem 2.13, we show that for the constraint from Lemma 4.7 which accepts all monomials of total degree at most $d$ while rejecting all the canonical monomials of degree $b_i(d)$, its orbit must accept all monomials $x^{\overline{d}}$ for $\overline{d} \in \text{Deg}(\text{RM}[n, d, q])$ while rejecting $x^{\overline{b}}$ for $\overline{b} \in \text{Border}(\text{RM}[n, d, q])$ and thus satisfies the conditions of Theorem 2.13.

We start by stating a simple property of canonical monomials.

**Lemma 4.8.** *Let $m$ be a monomial of total degree $d$, and let $\mathcal{F}$ be the minimal linear affine-invariant family containing $m$. Then $\mathcal{F}$ contains the canonical monomial of degree $d$ over $\mathbb{F}_q$.*

For the proof of the above lemma we shall need the following lemma.

**Lemma 4.9.** *Let $q = p^s$ for a prime $p$, and let $m = x_1^{d_1} x_2^{d_2}$ be a monomial such that $0 \le d_1 \le q-1$, $0 \le d_2 \le q - 1$ and $k \le_p d_2$. Let $\mathcal{F}$ be the minimal affine-invariant linear family containing $m$. Then the monomial $m' = x_1^{d_1+k} x_2^{d_2-k}$ is contained in $\mathcal{F}$.*

*Proof of Lemma 4.9.* Let $T$ be the affine-transformation $T(x_1, x_2) = (x_1, x_1 + x_2)$. Then

$$m \circ T = x_1^{d_1}(x_1 + x_2)^{d_2} = x_1^{d_1}\left(\sum_{i=0}^{d_2} \binom{d_2}{i} x_1^i x_2^{d_2-i}\right) = \sum_{i=0}^{d_2} \binom{d_2}{i} x_1^{d_1+i} x_2^{d_2-i}.$$

From Lucas's Theorem (Theorem 2.8) we have that $\binom{d_2}{i} \not\equiv 0 \mod p$ if and only if $i \leq_p d_2$. Thus we have that all monomials of the form $x_1^{d_1+i} x_2^{d_2-i}$ such that $i \leq_p d_2$ are contained in the support of $m \circ T$. Since $k \leq_p d_2$ we have that $m \circ T$ contains the monomial $m'$ in its support. Since $\mathcal{F}$ is affine-invariant we have that $m \circ T$ is contained in $\mathcal{F}$ and hence the monomial extraction lemma (Lemma 2.5) implies that $m'$ is contained in $\mathcal{F}$. □

Next we prove Lemma 4.8 based on Lemma 4.9.

*Proof of Lemma 4.8.* Write $d = \ell(q - q/p) + r'$ where $r' < q - q/p$. Let $m = \prod_{i=1}^{n} x_i^{d_i}$. We start by showing that $\mathcal{F}$ contains a monomial $\prod_{i=1}^{n} x_i^{d'_i}$ which satisfies $d'_i \geq q - q/p$ for every $2 \leq i \leq \ell + 1$.

We shall apply lemma 4.9 iteratively. For a monomial $m' = \prod_{i=1}^{n} x_i^{e_i}$ let $c(m') = \sum_{i=2}^{\ell+1} \min\{0, (q - q/p) - e_i\}$. Clearly, $c(m') = 0$ if and only if $e_i \geq q - q/p$ for all $2 \leq i \leq \ell + 1$. If $m$ satisfies that $d_i \geq q - q/p$ for all $2 \leq i \leq \ell + 1$ then we are done, hence assume that there exists $2 \leq i \leq \ell + 1$ such that $d_i < q - q/p$. If there exists a degree $j \in \{1\} \cup \{\ell + 2, \ldots, n\}$ such that $d_j > 0$ let $p^k$ be such that $p^k \leq_p d_j$. Lemma 4.9 implies that $m_1 := m \cdot x_i^{p^k} x_j^{-p^k}$ is contained in $\mathcal{F}$. Otherwise, since $\sum_{i=1}^{n} d_i = d$ there exists $j \in \{2, \ldots, \ell + 1\}$ such that $d_j > q - q/p$. Note that the base-$p$ representation of $q - q/p$ is $(q - q/p) = (p - 1) \cdot p^{s-1}$ and hence the fact that $d_j > q - q/p$ implies the existence of $p^k \leq_p d_j$ such that $q - q/p + p^k \leq_p d_j$. Lemma 4.9 implies that $m_1 := m \cdot x_i^{p^k} x_j^{-p^k}$ is contained in $\mathcal{F}$ in this case as well. Note that in both cases we have that $c(m_1) < c(m)$. Also, since $d_i < q - q/p$ we have that $d_i + p^k < q - q/p + q/p < q$. Hence all variables in $m_1$ have degree at most $q - 1$ (we mention this fact since this will allow us to apply Lemma 4.9 iteratively).

If all variables $x_2, \ldots, x_{\ell+1}$ in $m_1$ have degree at least $q - q/p$ then we are done. Otherwise repeat the same process as previously to obtain a monomial $m_2 \in \mathcal{F}$ such that the degrees of all variables in $m_2$ are at most $q - 1$ and $c(m_2) < c(m_1)$. Continuing this way we have that at the $i$-th step either all variables $x_2, \ldots, x_{\ell+1}$ in $m_{i-1}$ have degree at least $q - q/p$ and hence we are done or that we obtain a monomial $m_i \in \mathcal{F}$ such that the degrees of all variables in $m_i$ are at most $q - 1$ and $c(m_i) < c(m_{i-1})$. Since the function $c(m_i)$ strictly declines at each step the process must terminate eventually, and when it terminates we have that $m_{i-1}$ satisfies that all variables $x_2, \ldots, x_{\ell+1}$ in it have degree at least $q - q/p$.

We have just shown that $\mathcal{F}$ contains a monomial $m' = \prod_{i=1}^{n} x_i^{d'_i}$ where $d'_i \geq q - q/p$ for all $2 \leq i \leq \ell + 1$. Next we claim that the monomial $\tilde{m} = \prod_{i=1}^{n} x_i^{\tilde{d}_i}$ which satisfies $\tilde{d}_1 = r'$, $\tilde{d}_i = q - q/p$ for all $2 \leq i \leq \ell + 1$ and $\tilde{d}_i = 0$ for all $\ell + 2 \leq i \leq n$ is contained in $\mathcal{F}$ (note that $\tilde{m}$ is the canonical monomial of degree $d$ if and only if $r' + q - q/p > q - 1$). To see this note that if $m' = \tilde{m}$ then we are done. Otherwise we must have that $d'_1 < r'$. As was the case previously this implies the existence of either $\ell + 2 \leq j \leq n$ and an integer $k$ such that $p^k \leq_p d'_j$ or $2 \leq j \leq n$ such that $q - q/p + p^k \leq_p d'_j$. Lemma 4.9 implies that the monomial $m' \cdot x_1^{p^k} x_j^{-p^k}$ is contained in $\mathcal{F}$. Note that in the latter monomial the degree of the variable $x_1$ increased, but all variables $x_2, \ldots, x_{\ell+1}$ are still of degree at least $q - q/p$. Continuing this way we conclude that the monomial $\tilde{m}$ is contained in $\mathcal{F}$.

Finally, note that if $r' + q - q/p > q - 1$ then $\tilde{m}$ is also the canonical monomial of degree $d$ and hence we are done. Otherwise we have that $q - q/p \leq_p \tilde{d}_{\ell+1}$ and hence Lemma 4.9 implies that the monomial $\tilde{m} \cdot x_1^{q-q/p} \cdot x_{\ell+1}^{-(q-q/p)}$ is contained in $\mathcal{F}$. The proof is completed by noting that in this case the latter monomial is the canonical monomial of degree $d$.

□

**Lemma 4.10.** *For prime power $q = p^s$ and integers $n, d$, let $C$ be a constraint on $\{\mathbb{F}_q^n \to \mathbb{F}_q\}$ that accepts all monomials of total degree at most $d$ and rejects the canonical monomials of degree $b_i(d)$ for all $0 \leq i \leq s$. Then the orbit of $C$ accepts all monomials $x^{\overline{d}}$ with $\overline{d} \in \mathrm{Deg}(\mathrm{RM}[n, d, q])$ and rejects all monomials $x^{\overline{b}}$ with $\overline{b} \in \mathrm{Border}(\mathrm{RM}[n, d, q])$.*

*Proof.* The fact that $C$ accepts all monomials $x^{\overline{d}}$ with $\overline{d} \in \mathrm{Deg}(\mathrm{RM}[n, d, q])$ is syntactically equivalent to saying $C$ accepts all monomials of total degree at most $d$ and hence $C$ accepts all functions in $\mathrm{RM}[n, d, q]$. Since $\mathrm{RM}[n, d, q]$ is affine-invariant this implies in turn that the orbit of $C$ accepts all functions in $\mathrm{RM}[n, d, q]$ as well.

It just remains to show that the orbit of $C$ rejects every monomial $m = x^{\overline{b}}$ with $\overline{b} \in \mathrm{Border}(\mathrm{RM}[n, d, q])$. Assume for contradiction that the orbit of $C$ accepts $m$, and note that from Proposition 2.12 we have that $m$ is of total degree $b_i(d)$ for some $0 \leq i \leq s$.

Let $\mathcal{F}'$ be the set of functions accepted by the orbit of $C$, i.e.,

$$\mathcal{F}' = \{f : \mathbb{F}_q^n \to \mathbb{F}_q | T \circ C \text{ accepts } f \text{ for every affine-transformation } T\}.$$

We have that $\mathcal{F}'$ is linear and affine-invariant, and contains $m$, and so by Lemma 4.8 it also contains the canonical monomial of degree $b_i(d)$. So the orbit of $C$ accepts the canonical monomial of degree $b_i(d)$ contradicting the hypothesis about $C$. □

*Proof of Theorem 2.13.* From Lemma 4.7 we have a constraint $C$ of arity $k \leq 3q^4 \cdot \left(2^{p-1} + p - 1\right)^{(d+1)/(q(p-1))} \cdot q^{(d+1)/q}$ that accepts every monomial of total degree at most $d$ and rejects every canonical monomial of degree $b_i(d)$. By Lemma 4.10, $C$ satisfies the conditions necessary required in the theorem statement. □

## Acknowledgements

We would like to thank Amir Shpilka for suggesting that our tests are related to directional derivatives.